\documentstyle[prl,aps,psfig,floats,twocolumn]{revtex}

\begin{document}

\title{
\begin{flushright}
\small{LBNL-42184       \\
D\O\ Note 3487}  \\
\end{flushright}
\vskip 0.2 cm
W BOSON PHYSICS AT THE FERMILAB TEVATRON COLLIDER\thanks{
Invited plenary talk at the XVIII International Conference on
Physics in Collision, Frascati, Italy, June 17-19, 1998.}}

\author{Ronald~J.~Madaras}

\address{Lawrence Berkeley National Laboratory     \\
University of California, Berkeley, CA 94720, USA  \\
\vskip 0.3 cm
{\rm (For the CDF and D\O\ Collaborations)}
\vskip 0.8 cm
{\rm Recent results from the CDF and D\O\ Experiments at the Fermilab Tevatron
Collider are presented for the $W$ and $Z$ boson production cross sections,
the $W$ boson width, rare $W$ boson decays, trilinear gauge boson couplings,
and the $W$ boson mass.}}

\maketitle

\section{Introduction}

The CDF\cite{CDF_NIM} and D\O\cite{D0_NIM} detectors at the Fermilab 
Tevatron Collider collected data
during 1992-96 corresponding to an integrated luminosity of about 130 pb$^{-1}$
for each experiment.  This ``Run 1" was divided into three parts:

\vskip 0.1 cm
\begin{tabular}{lll}
\hskip 0.7 cm Run 1A   &1992--93     &$\sim$20 pb$^{-1}$ of luminosity\\
\hskip 0.7 cm Run 1B   &1994--95     &$\sim$90 pb$^{-1}$ \\
\hskip 0.7 cm Run 1C   &1995--96     &$\sim$20 pb$^{-1}$ \\
\end{tabular}
\vskip 0.1 cm

\noindent The large number of $W$ bosons detected (about 70,000 in the
$W\rightarrow e\nu$ channel by each experiment in Run 1A+B) permits one to 
make precise measurements of its properties.

\section{W and Z Production Cross Sections}

\begin{figure}
\centerline{\psfig{bbllx=30pt,bblly=251pt,bburx=579pt,bbury=542pt,figure=
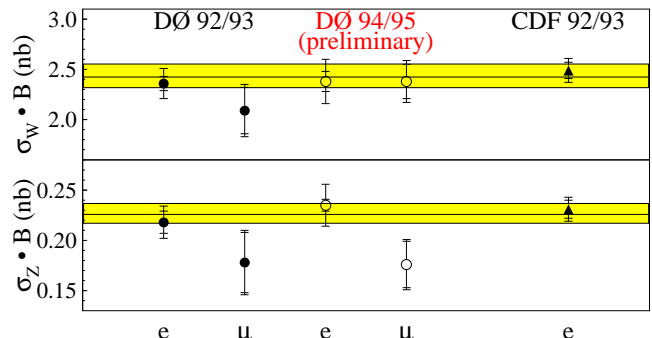,width=86mm}}
\caption{\it $\sigma \cdot B$ for inclusive $W$ and $Z$ production. The shaded
bands are the ${\cal{O}}(\alpha_s^2)$ theoretical QCD prediction.}
\label{WZ_XS}
\end{figure}

The measurement of the production cross sections times leptonic branching 
ratios ($\sigma \cdot B$) for $W$ and $Z$ bosons can be used to test QCD
predictions of $W$ and $Z$ boson production. The $W$ and $Z$ bosons are
detected via their leptonic decays: $W\rightarrow e\nu, \mu\nu, \tau\nu$ and
$Z\rightarrow ee, \mu\mu$.  For the $e$ and $\mu$ channels one selects $W$ 
events with one isolated high transverse momentum lepton ($p_T > 20-25$ GeV/c) 
and large missing transverse energy ({\mbox{$\not\!\!E_T$}} $> 20-25$ GeV), 
and $Z$ events with two
isolated leptons with $p_T > 20-25$ GeV/c.  The backgrounds are mainly due to
QCD and cosmic rays, and are typically $<15\%$ for the $W$ sample and 
$<5\%$ for the $Z$ sample.  
Recent results for CDF\cite{CDF_XS} and D\O\ are shown in 
Fig.\ref{WZ_XS}, and are compared to the ${\cal{O}}(\alpha_s^2)$ theoretical 
QCD prediction\cite{XS_TH}. One sees that there is excellent agreement, 
providing an important verification of QCD.

\begin{figure}
\centerline{\psfig{bbllx=0pt,bblly=0pt,bburx=567pt,bbury=567pt,figure=
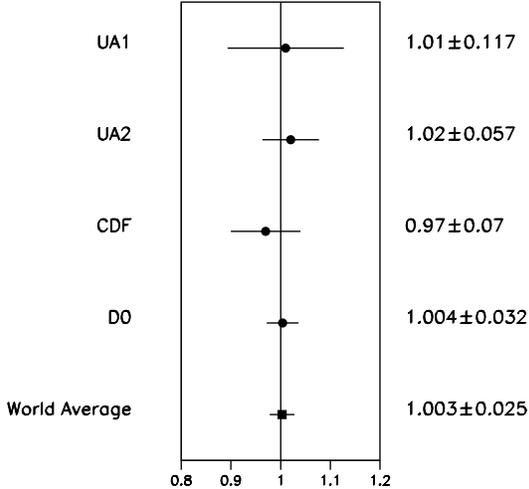,width=81mm}}
\caption{\it ${g_\tau^W}/{g_e^W}$ from various experiments.}
\label{TAU}
\end{figure}

D\O\ has also measured $W$ production cross sections by detecting 
$W\rightarrow \tau\nu$. The $\tau$ is identified through its hadronic decay
products, which are highly boosted and form a very narrow hadronic jet in the
D\O\ calorimeter. Thus one selects events with an isolated, narrow jet with
$E_T > 25$ GeV, and {\mbox{$\not\!\!E_T$}} $> 25$ GeV.  The {\it Profile}
variable, defined as the sum of the two highest $E_T$ towers divided by the 
$E_T$ of the jet, exploits the fine segmentation and good energy resolution of 
the D\O\ calorimeters to provide a powerful discrimination against QCD 
backgrounds. $W\rightarrow \tau\nu$ hadronic decays produce very narrow jets, 
leading to high values of {\it Profile}, and QCD jets yield wider jets, and 
therefore lower values of {\it Profile}. Events are selected with 
{\it Profile} $> 0.55$.  In a data sample of 17 pb$^{-1}$ D\O\ finds 1,202
candidate events, with a background of $222 \pm 16$ events.  The acceptance
$\times$ efficiency is $3.8\%$. The preliminary cross section times branching
ratio that D\O\ obtains is 
\begin{eqnarray}
{\sigma_W \cdot B(W\rightarrow \tau\nu) = 
2.38 \pm 0.09 \pm 0.10 \pm 0.20 ~{\rm nb,}}\nonumber
\end{eqnarray}
\noindent 
where the errors are statistical, systematic and luminosity, respectively.
Comparing this result with D\O's published value\cite{D0_XS} for 
$\sigma_W \cdot B(W\rightarrow e\nu)$ measures the ratio of the tau and electron
electroweak charged current couplings to the $W$ boson:
$$
{g_\tau^W}/{g_e^W} = 1.004 \pm 0.019 {\rm (stat)} \pm 0.026 {\rm (syst).}
$$
This result, shown in Fig.\ref{TAU} with the results of other experiments,
is in excellent agreement with $e-\tau$ universality.

\section{W Boson Width}

\subsection{Indirect Measurement of $\Gamma(W)$}

\begin{figure}[t]
\centerline{\psfig{bbllx=141pt,bblly=149pt,bburx=471pt,bbury=643pt,figure=
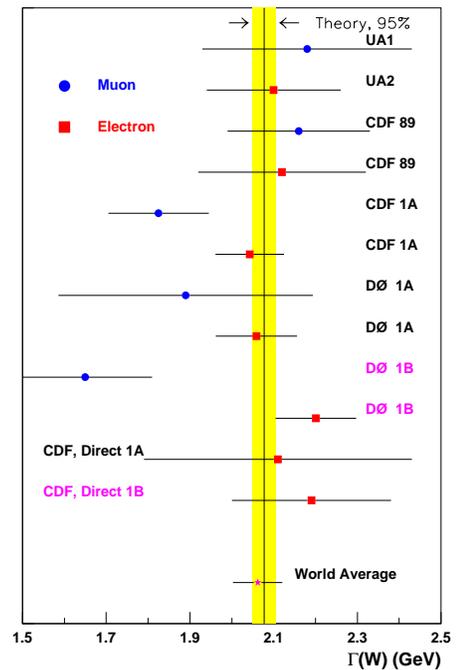,height=3.5in}}
\caption{\it Summary of W width measurements and comparison with the Standard
Model prediction.}
\label{SUM_GAMMA_W}
\end{figure}

One can indirectly measure the $W$ boson width from the ratio of the $W$ and
$Z$ production cross sections:
\begin{eqnarray}
R \ &&\equiv \ {{\sigma_W \cdot B(W\rightarrow l\nu)}\over
{\sigma_Z \cdot B(Z\rightarrow ll)}} \nonumber\\
 \ &&= \ {\left[\sigma_W \over \sigma_Z \right]}
 \cdot {1 \over B(Z\rightarrow ll)} \cdot {\Gamma(W\rightarrow l\nu) \over
 \Gamma(W)}
\label{Reqn}
\end{eqnarray}
where $l=e$ or $\mu$, $\sigma_W$ and $\sigma_Z$ are the inclusive
cross sections for $W$ and $Z$ boson production in {\mbox{$p\bar p$}}\
collisions, $B(W\rightarrow l\nu) = {\Gamma(W\rightarrow l\nu)}/\Gamma(W)$ is
the leptonic branching ratio of the $W$ boson,
and $B(Z\rightarrow ll)$ is the leptonic branching ratio of the $Z$ boson.
Many common sources of error cancel in $R$, including the uncertainty in the
luminosity and some of the errors in the acceptance and efficiency.
We extract the $W$ boson total width, $\Gamma(W)$, from Eq.\ref{Reqn} by
using the measured value of $R$, a theoretical calculation of 
$\sigma_W/ \sigma_Z$, the precise measurement of $B(Z\rightarrow ll)$
from LEP, and a theoretical calculation of ${\Gamma(W\rightarrow l\nu)}$.
Figure~\ref{SUM_GAMMA_W} summarizes $\Gamma(W)$ measurements 
from various experiments, along with the Standard Model prediction.  The
agreement of the experimental values with the theoretical prediction can be
used to set limits\cite{D0_XS} on unexpected decay modes of the $W$ boson, 
such as $W$ decays into supersymmetric charginos and neutralinos, or into heavy
quarks.

\subsection{Direct Measurement of $\Gamma(W)$}

\begin{figure}
\centerline{\psfig{bbllx=21pt,bblly=150pt,bburx=537pt,bbury=674pt,figure=
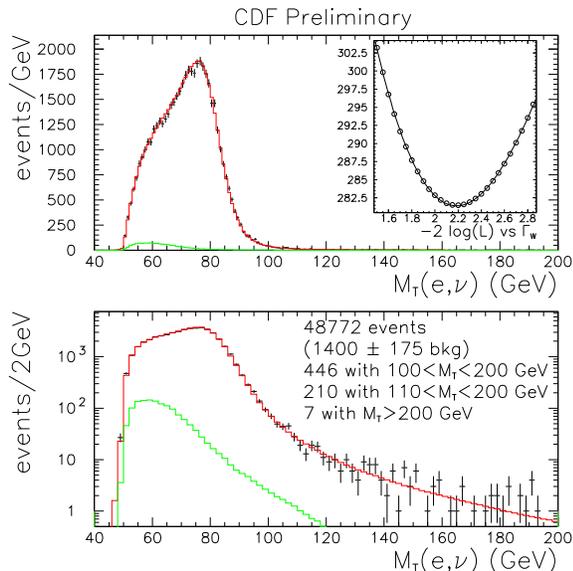,width=76mm}}
\caption{\it CDF $M_T$ distribution for $W \rightarrow e \nu$ events, with the 
best fit for $\Gamma_W$ 
overlayed. The size and shape of the background are also shown.}
\label{CDF_GAMMA_W}
\end{figure}

CDF has made a direct measurement of $\Gamma(W)$ from the 
$W$ boson transverse mass lineshape for $W \rightarrow e \nu$ events:
\begin{equation}
M_{T}^{2} = 2E_{T}^{l} E_{T}^{\nu} (1-cos \phi^{l \nu})
\end{equation}
A larger value of $\Gamma(W)$ increases the high transverse mass tail.
CDF determines
$\Gamma(W)$ from a binned likelihood fit to the $M_T$ spectrum in the region
$M_T > 110$ GeV/c$^2$, where the Breit-Wigner line shape dominates over the
Gaussian resolution of the detector.  The CDF $M_T$ spectrum and fit are
shown in Fig.\ref{CDF_GAMMA_W}, and the preliminary value from this analysis
of Run 1B data is:
\begin{eqnarray}
{\Gamma(W) = 2.19^{+0.17}_{-0.16} {\rm (stat)} \pm{0.09} {\rm (syst)}
{\rm~GeV.}}\nonumber
\end{eqnarray}
This result is in good agreement with the indirect measurements and the 
SM prediction.

\begin{figure}[t]
\centerline{\psfig{bbllx=12pt,bblly=154pt,bburx=536pt,bbury=654pt,figure=
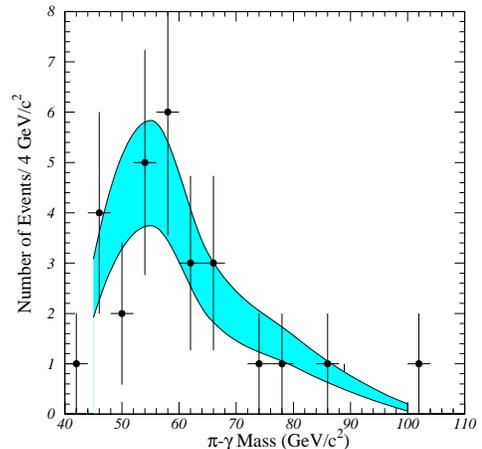,height=60mm}}
\caption{\it Distribution of the $\pi-\gamma$ mass for the 28 CDF 
$W \rightarrow \pi \gamma$ candidates.
The shaded band shows the one sigma uncertainty in the background expectation
value.}
\label{PIGAMMA}
\end{figure}

\section{Rare W Decays}

\subsection{$W\rightarrow \pi \gamma$}

The ratio of the partial widths of the decays $W\rightarrow \pi \gamma$ to
$W\rightarrow e \nu$ is predicted\cite{WDECAY_THEOR} to be 
$\Gamma(W\rightarrow \pi \gamma)/\Gamma(W\rightarrow e \nu) \simeq 3 \times
10^{-8}$.  CDF\cite{PIGAMMA_OLD_CDF} has the best previous
experimental limit on this ratio of $2.0 \times 10^{-3}$.
CDF now has new results\cite{PIGAMMA_NEW_CDF} 
on $W\rightarrow \pi \gamma$ based
on 83 pb$^{-1}$ of data taken in Run 1B (1994-95).  They chose events with one
isolated photon with $p_T > 23$ GeV/c and one jet consistent with a single,
isolated charged pion with $p_T > 15$ GeV/c, separated by $\Delta \phi > 1.5$
radians, and no other jets with $E_T >$ 15 GeV.  The $\pi-\gamma$ masses of
the 28 events that result from these cuts are shown in Fig.\ref{PIGAMMA}, along
with the estimate of the background (which is due mainly to QCD direct
photons).  There are 3 events in the $W$ mass region, with an estimated
background of $5.2 \pm 1.5$ events.  The acceptance $\times$ efficiency is
$3.8\%$.  Thus CDF finds at the $95\%$ CL that $\sigma_W \cdot B(W\rightarrow 
\pi \gamma) < 1.7$ pb, and that $\Gamma(W\rightarrow \pi \gamma)/\Gamma(W
\rightarrow e \nu) < 7 \times 10^{-4}$.  This limit is a factor of three times
better than their previous limit.

\subsection{$W\rightarrow D_s \gamma$}

The theoretical prediction\cite{WDECAY_THEOR} for the ratio of the partial
widths
$\Gamma(W\rightarrow D_s \gamma)/\Gamma(W\rightarrow e \nu)$ is $1 \times
10^{-7}$, which is three times larger than the relative branching fraction for
the decay $W\rightarrow \pi \gamma$.  However, the multitude of $D_s$ decay
modes and the choice of particular modes for experimental identification makes
the experimental reach smaller in the $W\rightarrow D_s \gamma$ case.  CDF has
put a limit\cite{DS_GAMMMA_CDF} on this relative branching fraction using 
82 pb$^{-1}$ of data from Run 1B (1994-95). They select events with one 
isolated photon with $p_T > 22$ GeV/c, and 
one isolated $D_s$ candidate with $p_T > 22$ GeV/c. The $D_s$ mesons are
identified via the decay modes $D_s \rightarrow \phi \pi$ (with $\phi
\rightarrow KK$), and $D_s \rightarrow K^{*0}K$ (with $K^{*0} 
\rightarrow K \pi$).  They find 4 events with a $D_s-\gamma$ mass consistent
with the $W$, with an estimated background of 4 events (mainly due to QCD
direct photons). The acceptance $\times$ efficiency is $6.9\%$.  
Thus CDF finds at the $95\%$ CL that $\sigma_W \cdot B(W\rightarrow 
D_s \gamma) < 27.4$ pb, and that $\Gamma(W\rightarrow D_s \gamma)/\Gamma(W
\rightarrow e \nu) < 1.1 \times 10^{-2}$.  This is the first measurement of
this quantity.

\section{Trilinear Gauge Boson Couplings}

\begin{figure}[h]
\centerline{\psfig{bbllx=68pt,bblly=271pt,bburx=548pt,bbury=520pt,figure=
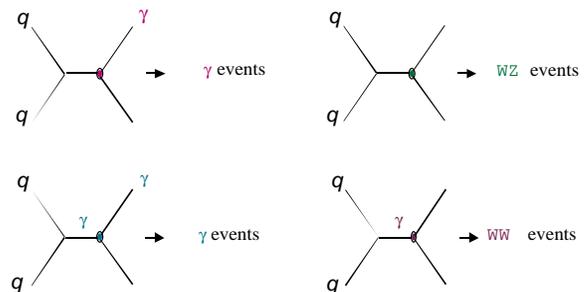,width=76mm}}
\caption{\it Measurement of the trilinear gauge boson couplings
$WW\gamma$, $WWZ$, $ZZ\gamma$, and $Z\gamma\gamma$ using diboson events.}
\label{DIBOSON}
\end{figure}

The Standard Model (SM) predicts the existence of gauge boson self-interactions,
and makes unique predictions for the strength of these trilinear gauge boson
couplings.  Measurements of these couplings test the SM, and any significant
deviation from SM predictions would be compelling evidence for new physics.
As is seen in Fig.\ref{DIBOSON},
the direct measurement of these trilinear couplings ($WW\gamma, WWZ, ZZ\gamma,
{\rm and} Z\gamma\gamma$) is possible by measuring diboson production at the
Tevatron\cite{ELLISON}.

$WWV$ ($V=\gamma$ or $Z$) couplings are characterized by the parameters
$\Delta\kappa_V(\equiv\kappa_V-1)$ and $\lambda_V$,
which are equal to zero in the SM.  $ZV\gamma$ ($V=\gamma$ or $Z$)
couplings are characterized by $h^V_{30}$ and $h^V_{40}$, which are also zero in
the SM. To obey unitarity, all couplings are multiplied by a form factor
$(1+\hat{s}/\Lambda^2)^n$, where n=2 for $WWV$ couplings, 3 for $h^V_{30}$, and
4 for $h^V_{40}$, $\hat{s}$ is the square of the sub process center-of-mass 
energy, and $\Lambda$ is the form factor scale.  Anomalous (i.e. non Standard 
Model) values of the coupling parameters increase the diboson production cross 
section and enhance the $p_T$ spectrum of the gauge bosons for large values of 
$p_T$.

\subsection{$W\gamma$ Production}

\begin{figure}
\centerline{\psfig{bbllx=65pt,bblly=175pt,bburx=570pt,bbury=675pt,figure=
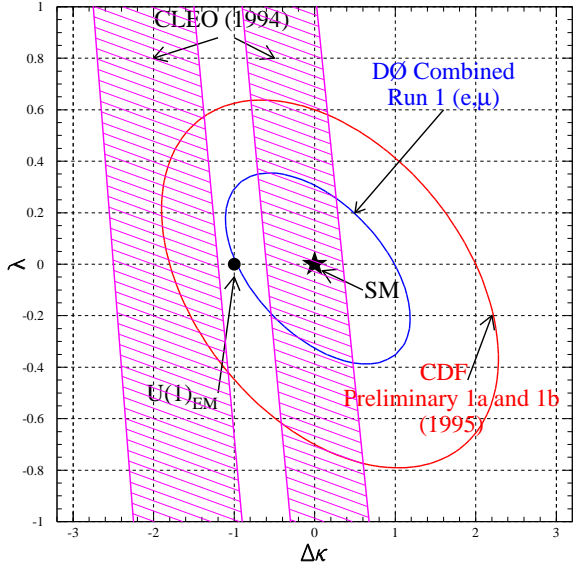,width=76mm}}
\caption{\it Contour limits on anomalous $WW\gamma$ couplings, from
$W\gamma$ events.}
\label{WGAMMA}
\end{figure}

The detection of $W\gamma$ events enables one to measure the $\lambda_\gamma$
and $\Delta \kappa_\gamma$ parameters that characterize $WW\gamma$ couplings.
One uses the leptonic decays of the $W$, and selects events with an isolated
high $p_T$ muon or electron, and with large {\mbox{$\not\!\!E_T$}}. The event
must also have an isolated photon with $E_T >$ 10 GeV (D\O) or 7 GeV (CDF).
The main background is $W$+jets, where the jet fragments into a $\pi^0$, and
$\pi^0 \rightarrow \gamma\gamma$.  From the Run 1B data set, D\O\cite{D0_WGAMMA}
finds 127 candidate events with 93 pb$^{-1}$, and CDF\cite{CDF_WGAMMA} finds 
109 events with 67 pb$^{-1}$. The anomalous coupling parameters are determined 
from a binned likelihood fit to the $p_T(\gamma)$ spectrum, and are shown in 
Fig.\ref{WGAMMA}.
The D\O\ limits at $95\%$ CL, for $\Lambda=1.5$ TeV, are $-0.93 < 
\Delta\kappa_\gamma < 0.94$ (for $\lambda_\gamma=0$), and $-0.31 <
\lambda_\gamma < 0.29$ (for $\Delta\kappa_\gamma=0$).  These limits are
independent of the $WWZ$ vertex, unlike the limits obtained from $WW$
production. From Fig.\ref{WGAMMA} one sees that the D\O\ results
exclude the coupling $\lambda_\gamma = \kappa_\gamma = 0$
at the $95\%$ CL, providing the first
direct evidence that the photon couples to more than just the electric charge of
the $W$ boson.

\subsection{$WW \rightarrow l\nu l\nu~~(l=e,\mu)$}

These events are selected by requiring two isolated leptons with $p_T > 15-25$
GeV/c, and {\mbox{$\not\!\!E_T$}} $> 20-25$ GeV.  The main backgrounds are
due to $t\bar{t}, Z \rightarrow \tau\tau,$ and Drell-Yan production.  In a
97 pb$^{-1}$ sample D\O\cite{D0_WW_LL} finds 5 events, with a background of
$3.1 \pm 0.4$ events, and sets a 95$\%$ CL upper limit on $\sigma_{WW}$ of
37.1 pb.  In a 108 pb$^{-1}$ sample CDF\cite{CDF_WW_LL} also finds 5 events, 
but with
a lower background of $1.2 \pm 0.3$ events, and thus measures $\sigma_{WW} =
10.2^{+6.3}_{-5.1} \pm 1.6$ pb.  The SM prediction is $\sigma_{WW} =
9.5 \pm 1.0$ pb, so there is no evidence for anomalous WW production.  To get
limits on the anomalous coupling parameters, CDF fits to the total number of
events.  D\O\ fits to the lepton $p_T$ spectrum, which gives significantly
better limits.
D\O\ finds, for $\Lambda=1.5$ TeV, and assuming $\Delta\kappa_Z=\Delta\kappa_
\gamma$ and $\lambda_Z=\lambda_\gamma$:
$-0.62 < \Delta\kappa < 0.77$ (for $\lambda=0$), and 
$-0.53 < \lambda < 0.56$ (for $\Delta\kappa=0$).

\begin{figure}
\centerline{\psfig{bbllx=60pt,bblly=163pt,bburx=552pt,bbury=634pt,figure=
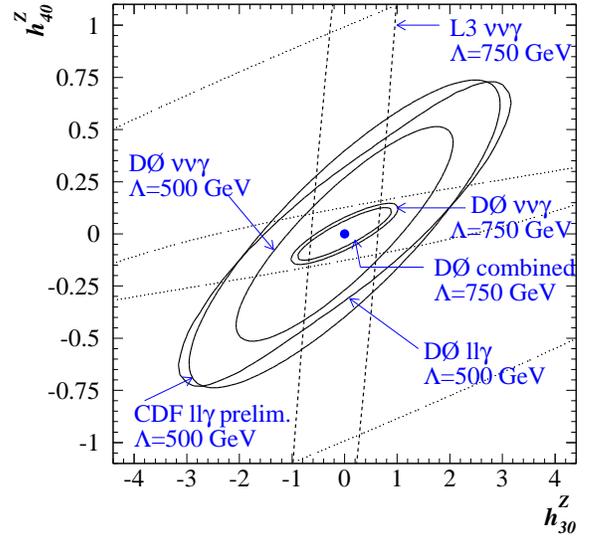,width=76mm}}
\caption{\it Contour limits on anomalous $ZZ\gamma$ couplings, from
$Z\gamma$ events.}
\label{ZGAMMA}
\end{figure}

\subsection{$WW, WZ \rightarrow l\nu jj, lljj~~(l=e,\mu)$}

These events are selected by requiring one isolated lepton with $p_T > 20-25$
GeV/c, two or more jets with $E_T > 20-30$ GeV which have an invariant mass 
consistent
with a $W$ or a $Z$, and {\mbox{$\not\!\!E_T$}} $> 20-25$ GeV (or a second
high $p_T$ lepton for the $lljj$ events).  The background from $W$+jets is large
in this channel.  CDF\cite{CDF_WW_LJETS} uses events with $p_T(jj) > 200$ GeV/c
to get anomalous coupling limits, and D\O\cite{D0_WW_LJETS} uses a binned
likelihood fit to the $p_T(W)$ spectrum to get them.  The limits on the
anomalous couplings $\Delta\kappa$ and $\lambda$ obtained by each experiment 
are similar, and are about a factor of 1.4 tighter than those from the 
$WW \rightarrow l\nu l\nu$ channel.  
The coupling $\lambda_Z = \kappa_Z = 0$ is excluded at $> 99\%$ CL by both
experiments, thus providing the first direct evidence for a $WWZ$ coupling.

\subsection{$Z\gamma$ Production}

D\O\cite{D0_ZGAMMA} and CDF\cite{CDF_WGAMMA} have each measured $Z(ee)\gamma$ 
and $Z(\mu\mu)\gamma$ production.
D\O\ (CDF) finds 35 (33) events in 105 (67) pb$^{-1}$, with a background of
5.9 (1.4) events. The measurements agree with Standard Model expectations, and
limits on the anomalous coupling parameters are found using a binned maximum
likelihood fit to the photon $E_T$ spectra.  The results are the outer two
ellipses in Fig.\ref{ZGAMMA}.

D\O\ has also measured\cite{D0_ZGAMMA_NN} $Z(\nu\nu)\gamma$ production.
The sensitivity to anomalous couplings is much higher in the $Z(\nu\nu)\gamma$
channel than in the $Z(ll)\gamma$ channel due to a higher branching ratio
and the absence of diluting radiative Z decay events. But the measurement of
$Z(\nu\nu)\gamma$ production is very challenging at a hadron collider because
of the extremely high background (due to muon bremsstrahlung,
$W\rightarrow e\nu$, jet-jet and jet-$\gamma$ production, etc.). Features of
D\O\ that enable them to do this measurement are:

\begin{figure}
\centerline{\psfig{bbllx=21pt,bblly=92pt,bburx=524pt,bbury=416pt,figure=
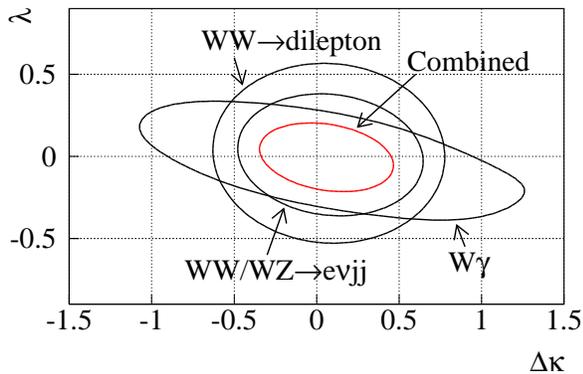,width=76mm}}
\caption{\it D\O\ contour limits on anomalous $WW\gamma$ couplings (for 
$\Lambda = 1.5$ TeV).}
\label{FOUR_CONTOURS}
\end{figure}

\vskip 0.2 cm
\noindent{Hermeticity:}
The excellent hermeticity of the D\O\ calorimeter results in a small tail in
the missing $E_T$ resolution, and reduces the QCD background.

\vskip 0.2 cm
\noindent{Hit Counting:}
Because of the high hit efficiency of the tracking chamber, one
can count hit wires to help eliminate background due to $W\rightarrow e\nu$,
even if the track for the electron is not reconstructed.

\vskip 0.2 cm
\noindent{Photon ``Tracking" in the Calorimeter:}
Because of the fine longitudinal and transverse segmentation in the D\O\
electromagnetic calorimeter, one can determine the direction of the photon and
determine if it came from the primary vertex, and thus reduce the
muon bremsstrahlung background from cosmics and beam halo.

\vskip 0.2 cm
\noindent{Muon ``Tracking" in the Calorimeter:}
Because one can detect minimum ionizing particles in the D\O\ calorimeter, one
can reduce the muon bremsstrahlung background from cosmic rays and beam halo
by searching for a line of minimum ionizing hits in the calorimeter.

\vskip 0.2 cm
\noindent In the $Z(\nu\nu)\gamma$ channel D\O\ finds 4 events, with a
background of $5.8 \pm 1.0$ events, for 13 pb$^{-1}$. One expects $1.8 \pm 0.2$
events from the Standard Model.  Anomalous coupling limits are found using a
binned maximum likelihood fit to the $E_T(\gamma)$ spectrum, and are shown as
the inner ellipses in Fig.\ref{ZGAMMA}.  Combining the results from the 
$Z(ll)\gamma$ and $Z(\nu\nu)\gamma$ channels, D\O\cite{D0_ZGAMMA} finds, for
$\Lambda=750$ GeV:
\begin{eqnarray}
\mid h_{30}^{Z,\gamma} \mid < 0.37 ~~{\rm and}~~
\mid h_{40}^{Z,\gamma} \mid < 0.05 \nonumber
\end{eqnarray}
\noindent These are the most stringent direct
limits on anomalous couplings from any experiment.

\begin{table}
\setlength{\tabcolsep}{5pt}
\caption{\it
D\O\ limits on anomalous couplings $\alpha_{B\phi}$,
$\alpha_{W\phi}$, and $\alpha_{W}$ at the 95\% CL from a simultaneous fit
to the $W\gamma, WW \rightarrow \ell\nu \ell\nu$, and $WW/WZ \rightarrow 
e\nu jj$ data.  Also shown are the LEP limits, and the LEP + D\O\ 
combined limits.}
\vskip 0.1 in
\begin{tabular}{cccc}  
& D\O\     & LEP      & D\O\ + LEP \\
\hline
$\alpha_{B\phi}$ & -0.77, 0.58 & -0.44, 0.95 & -0.42, 0.43 \\
$\alpha_{W\phi}$ & -0.22, 0.44 & -0.12, 0.13 & -0.14, 0.10 \\
$\alpha_W$       & -0.20, 0.20 & -0.21, 0.27 & -0.18, 0.13 \\
\end{tabular}
\label{ALPHA_COUPLINGS}
\end{table}

\subsection{D\O\ Combined Analysis of $WW\gamma$ and $WWZ$ Couplings}

D\O\ has performed\cite{D0_COMB_WG} a simultaneous fit to the photon $p_T$
spectrum in the $W\gamma$ data, the lepton $p_T$ distribution in the $WW$
dilepton data, and the $W~p_T$ distribution in the $WW/WZ \rightarrow e \nu jj$
data.  The limits on the $WW\gamma$ and $WWZ$ anomalous coupling parameters are
extracted from the fit taking correlations properly into account, and are
shown in Fig.\ref{FOUR_CONTOURS}, assuming identical $WW\gamma$ and $WWZ$
couplings.  The $95\%$ CL limits, for $\Lambda=2.0$ TeV, are:
\begin{eqnarray}
-0.30 < \Delta\kappa < 0.43 ~~({\rm for} \lambda=0) \nonumber\\
-0.20 < \lambda < 0.20 ~~({\rm for} \Delta\kappa=0) \nonumber
\end{eqnarray}
\noindent The D\O\ simultaneous fit has
also been done using the alternative parameterization of the anomalous
couplings used by the LEP groups: $\alpha_{B\phi}, \alpha_{W\phi}$
and $\alpha_W$.
The resulting limits are listed in Table~\ref{ALPHA_COUPLINGS}. Also listed are 
the limits on the anomalous coupling parameters from combining\cite{LEP} the 
D\O\ and LEP results.  No anomalous diboson production has been seen at either 
the Tevatron or LEP, and stringent limits have been set on the anomalous 
coupling parameters.

\section{W Boson Mass}

The $W$ boson mass is a fundamental parameter of the Standard Model.  At next
to leading order it can be written as:
\begin{equation}
M_W^2 = {{\pi\alpha(M_Z)} \over {\sqrt{2}G_F}} {1 \over {(1-(M_W^2/M_Z^2))}}
{1 \over (1-\Delta r)}
\end{equation}
where M$_Z$ is the $Z$ boson mass, G$_F$ is the Fermi coupling constant,
$\alpha$ is the fine structure constant (evaluated at a scale=$M_Z$), and
$\Delta r$ represents the effect of radiative corrections.  The parameters
M$_Z$, G$_F$ and $\alpha$ have been measured to better than $0.01\%$.
The parameter $\Delta r$ depends on the masses of particles which couple to
the $W$ boson, such as the top quark, Higgs boson or new particles.  Thus a
precision measurement of the $W$ boson mass, along with a measurement of the
top quark mass, can be used to constrain the Higgs boson mass or indicate the
presence of new physics beyond the Standard Model.

\begin{figure}
\centerline{\psfig{bbllx=192pt,bblly=292pt,bburx=420pt,bbury=503pt,figure=
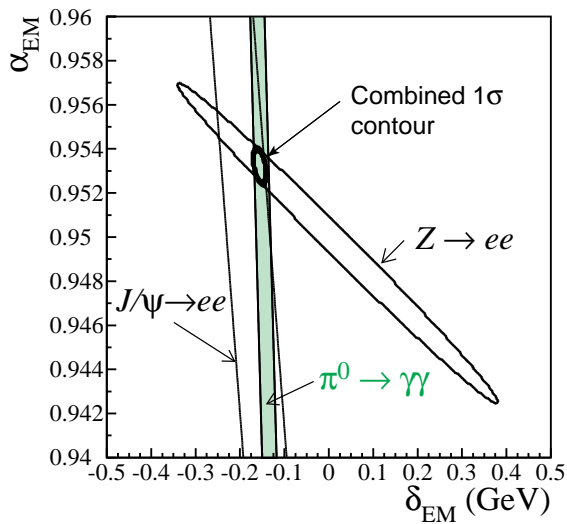,width=76mm}}
\caption{\it Constraints on the D\O\ electromagnetic energy scale parameters
from collider data.}
\label{D0_ESCALE}
\end{figure}

At the Fermilab Tevatron, $W$ bosons are produced via $p\bar{p} \rightarrow W$
+ jets, and the $W$ bosons are detected through their leptonic decays:
$W \rightarrow$ lepton + $\nu$. One can measure P$_{T}(\nu)$ from
transverse energy balance, but one can't measure P$_{L}(\nu)$ because of the
unknown amount of energy that went down the beampipe in the forward/backward
directions. Thus a true invariant mass cannot be calculated. Instead, one
calculates the ``transverse mass", as given in Eq.2.
The M$_T$ distribution shows a sharp Jacobian peak at the $W$ mass. The $W$
mass is determined from a likelihood fit of the M$_T$ distribution to Monte
Carlo generated templates in transverse mass for different $W$ mass values.
The E$_{T}^{\nu}$ measurement depends on the ``recoil" momentum of the
hadrons, and thus one needs to understand the resolution of, and bias in,
both the charged lepton energy measurement and the hadronic recoil
measurement in order to correctly model M$_T$ in the Monte Carlo.  These are
briefly discussed below.

\begin{figure}
\centerline{\psfig{bbllx=32pt,bblly=413pt,bburx=532pt,bbury=646pt,figure=
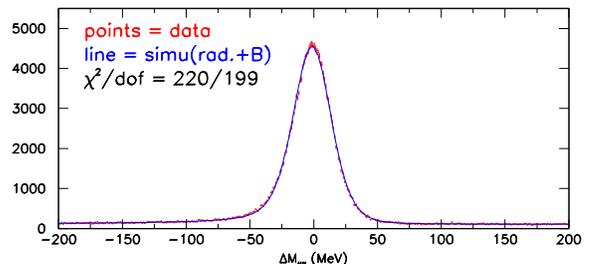,width=76mm}}
\caption{\it Difference between the CDF dimuon mass and the PDG value for
$J/\psi \rightarrow \mu\mu$ events.}
\label{CDF_PSCALE}
\end{figure}

The most recent Tevatron measurements of $M_W$, using the Run 1B (1994-95)
data, are D\O's published result\cite{D0_MW} using the $W \rightarrow e \nu$ 
channel, and CDF's preliminary result\cite{CDF_MW} using the $W \rightarrow
\mu \nu$ channel.  The experiments select events with an isolated, high
quality lepton in the central region with $p_T > 25$ GeV/c, 
{\mbox{$\not\!\!E_T$}} $> 25$ GeV, and hadronic recoil $< 15-20$ GeV.  This
results in a sample of 28K $W \rightarrow e \nu$ events for D\O, and 21K
$W \rightarrow \mu \nu$ events for CDF.

The D\O\ electromagnetic (EM) calorimeter energy scale is initially set by
test beam measurements, and then finally determined from collider data.  The 
observed EM energy is
parameterized as $E_{obs} = \alpha E_{true} + \delta$, and the parameters 
$\alpha$ and $\delta$ are determined from $Z \rightarrow ee, \pi^0 \rightarrow
\gamma\gamma$, and $J/\psi \rightarrow ee$ collider data, as shown in
Fig.\ref{D0_ESCALE}.  D\O\ finds that $\alpha = 0.9533 \pm 0.0008$ and
$\delta = -0.16_{-0.21}^{+0.03}$ GeV, including systematic errors from 
underlying event corrections and nonlinearity at low $E_T$.  The uncertainty on 
$\alpha$ ($\delta$) results in an error on $M_W$ of 65 (20) MeV.  D\O\ uses
its $Z \rightarrow ee$ sample to measure the constant term in the EM
energy resolution, and the measured uncertainty in the energy resolution results
in an error on $M_W$ of 20 MeV.

The momentum scale of the CDF central tracker is determined by normalizing the
observed $J/\psi \rightarrow \mu\mu$ peak to the world average, as is seen
in Fig.\ref{CDF_PSCALE}.  They find $\Delta M_{J/\psi}=0.7 \pm 1.5$ MeV.  The
uncertainty on $\Delta M_{J/\psi}$ results in an error on $M_W$ of 40 MeV.  CDF
uses its $Z \rightarrow \mu\mu$ sample to measure the momentum resolution, and
the measured uncertainty in the momentum resolution results in an error on 
$M_W$ of 25 MeV.

Both D\O\ and CDF use the transverse energy balance of the $Z$ boson and the
hadronic recoil products in $p\bar{p} \rightarrow Z + X$ events to determine
the hadronic recoil energy scale and resolution.  D\O\ uses $Z \rightarrow ee$
events, and CDF uses $Z \rightarrow \mu\mu$ events. Thus the hadronic recoil
scale is measured relative to the lepton energy scale.  The error on $M_W$ due
to the uncertainty in the hadron recoil scale and resolution, and the
uncertainty on the recoil model, is 35 (90) MeV
for D\O\ (CDF).  

\begin{figure}
\centerline{\psfig{bbllx=26pt,bblly=135pt,bburx=523pt,bbury=601pt,figure=
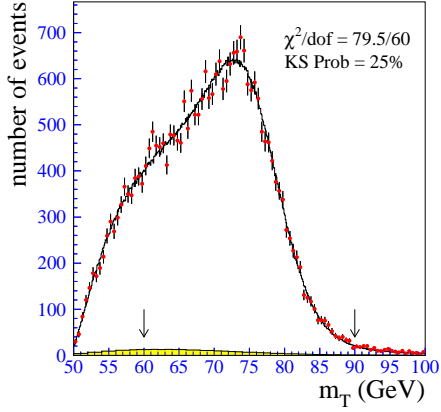,height=55mm}}
\caption{\it Transverse mass distribution of $W \rightarrow e\nu$ events from 
the D\O\ Run 1B data, with the best fit. The shaded distribution is the 
background.}
\label{D0_MW}
\end{figure}

\begin{figure}
\centerline{\psfig{bbllx=59pt,bblly=251pt,bburx=553pt,bbury=544pt,figure=
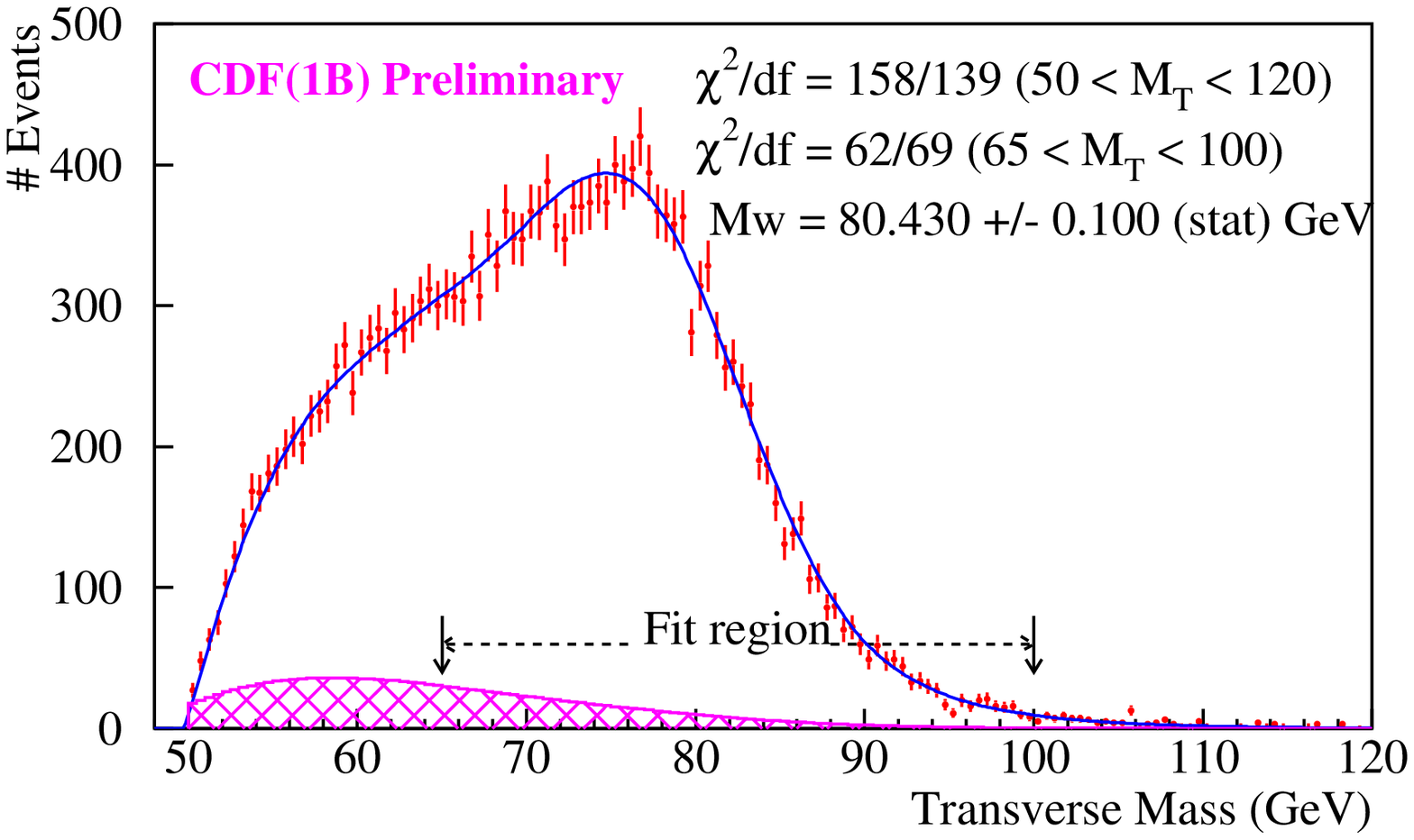,width=76mm}}
\caption{\it Transverse mass distribution of $W \rightarrow \mu\nu$ events from 
the CDF Run 1B data, with the best fit. The shaded distribution is the 
background.}
\label{CDF_MW}
\end{figure}

The fits to the $M_T$ distributions are shown in Fig.\ref{D0_MW} for D\O\ and 
in Fig.\ref{CDF_MW} for CDF. The results for these Run 1B measurements are:
\begin{eqnarray}
{\rm D\O~1B:}  \ &&M_W \ = \ 80.440 \pm 0.095 \pm 0.065 
{\rm~GeV/c}^2 \nonumber\\
{\rm CDF~1B:}  \ &&M_W \ = \ 80.430 \pm 0.100 \pm 0.120 
{\rm~GeV/c}^2 \nonumber
\end{eqnarray}
where the first error is statistical, and the second is systematic.
Table~\ref{MW_ERRORS} summarizes the sources of uncertainty in each of the
measurements. 
Combining these results with those of D\O\cite{D0_MW1A} and CDF\cite{CDF_MW1A}
from Run 1A gives:
\begin{eqnarray}
{\rm D\O~1A+B:}  \ &&M_W \ = \ 80.430 \pm 0.110 {\rm ~GeV/c}^2 \nonumber\\
{\rm CDF~1A+B:}  \ &&M_W \ = \ 80.375 \pm 0.120 {\rm ~GeV/c}^2 \nonumber
\end{eqnarray}
Combining these results with those of UA2\cite{UA2_MW} gives a Hadron Collider 
Average of:
\begin{eqnarray}
{\rm Hadron~Collider~Average:}~M_W =  80.400 \pm 0.090 {\rm ~GeV/c}^2 \nonumber
\end{eqnarray}

\begin{table}
\caption{\it Summary of Errors on $M_W$ (in MeV/$c^2$) for the
Run 1B Measurements.}
\vskip 0.1 in
\begin{tabular}{lcc}  
\\		& CDF 	& D\O\  \\
\underline{Statistical} 	&	& \\
~~W sample 			& 100	& 70 \\
~~Z sample (e energy scale) 	&\underline{~--~} &
\underline {~65~} \\
Total Statistical 		& 100	& 95 \\
\\
\underline{Systematic}\\
~~Muon momentum scale 		& 40 & -- \\
~~Lepton energy resolution 	& 25 & 20 \\
~~Calorimeter linearity 		& -- & 20 \\
~~Recoil modeling 		& 90 & 35 \\
~~W production model		& 55 & 30 \\
~~Backgrounds 			& 25 & 10 \\
~~Lepton angle calibration 	& -- & 30 \\
~~Fitting 			& 10 & -- \\
~~Miscellaneous 		&\underline{~15~} & 
\underline{~10~} \\
Total Systematic 		&\underline{\underline{120~}} 
& \underline{\underline{~65~}} \\
\\
\underline{Total Uncertainty} 		& 155 & 115 \\ 
\\
\end{tabular}
\label{MW_ERRORS}
\end{table}

\begin{figure}
\centerline{\psfig{bbllx=69pt,bblly=156pt,bburx=552pt,bbury=694pt,figure=
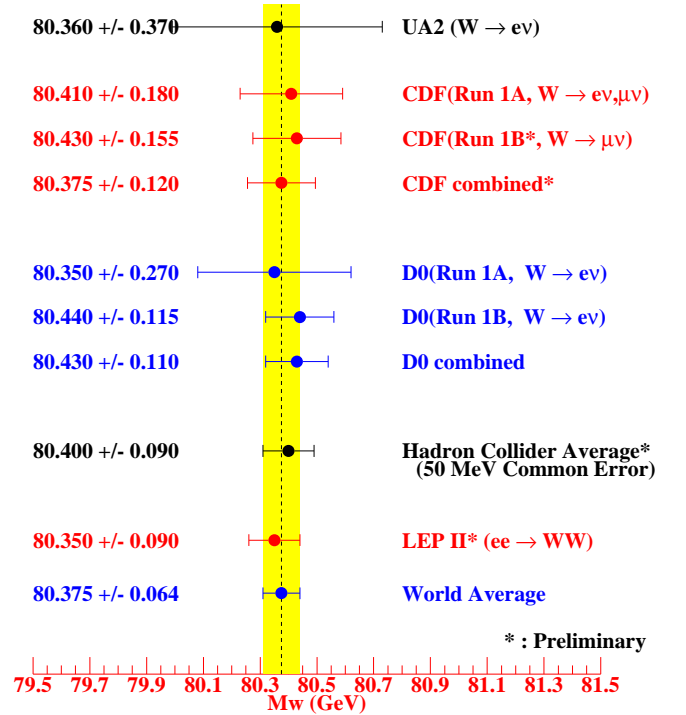,width=86mm}}
\caption{\it Summary of direct $W$ mass measurements.}
\label{DIRECT_MW}
\end{figure}

\begin{figure}
\centerline{\psfig{bbllx=1pt,bblly=23pt,bburx=503pt,bbury=495pt,figure=
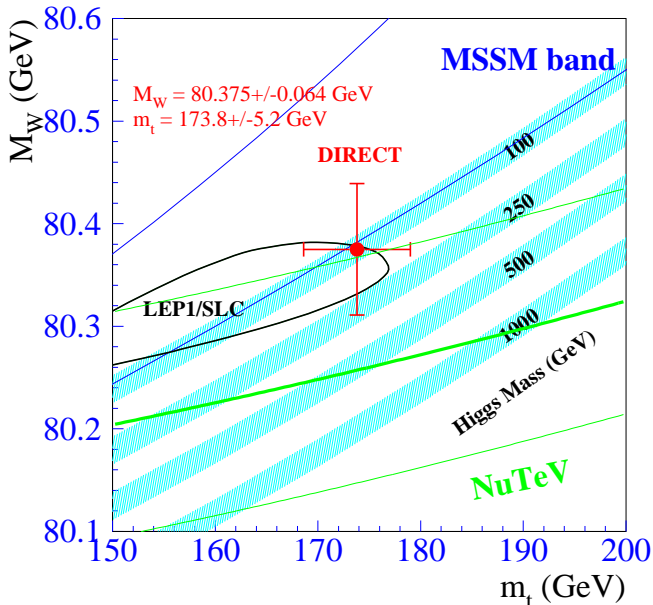,width=86mm}}
\caption{\it $M_W$ vs $M_{top}$. The point is the combined result from direct
measurements. Also shown are the allowed regions from LEP1/SLC and NuTeV, the
prediction of the minimal supersymmetric model (MSSM), and the Standard Model
predictions for Higgs masses from $100-1000$ GeV/c$^2$.}
\label{MW_MT_WORLD}
\end{figure}

\noindent Combining the Hadron Collider Average with the LEP2 
result\cite{LEP2} of
$M_W = 80.350 \pm 0.090$ GeV/c$^2$ presented at this conference gives a World
Average of direct $M_W$ measurements:
\begin{eqnarray}
{\rm Direct~World~Average:}~~M_W =  80.375 \pm 0.064 {\rm ~GeV/c}^2 \nonumber
\end{eqnarray}
These direct $M_W$ measurements are summarized in Fig.\ref{DIRECT_MW}. In
Fig.\ref{MW_MT_WORLD} $M_W$ is plotted versus $M_{top}$. The point is the 
Direct World Average, with $M_{top}$ taken from D\O\ and CDF measurements.  
Also shown are the indirect LEP1/SLC and NuTeV\cite{NUTEV} measurements, 
the prediction of the Minimal SuperSymmetric Model (assuming no SUSY particles
have masses low enough to be discovered at LEP2),
and the Standard Model predictions for Higgs masses from $100-1000$ GeV/c$^2$.

Most of the systematic errors in the Tevatron $M_W$ measurements are still
statistics limited, since they are determined with collider data. Thus
we expect improvements in both the short and long term future.  With the
Run 1 data, D\O\ is using its forward electrons, and expects to have a final
$\Delta M_W$ of less than 100 MeV.  CDF is finalizing its muon results with
smaller errors, and also using Run 1B electrons, and expects to have a final
$\Delta M_W$ of about 90 MeV.  Thus one expects a final Tevatron Run 1 
$\Delta M_W$ of about 75 MeV.  Run 2 at the Tevatron Collider, scheduled to
begin in April 2000, will have 20 times more integrated luminosity than
Run 1.  D\O\ is upgrading its tracking system, and adding new preshower
detectors and a new solenoid (which will enable them to also use muons to
measure $M_W$).  CDF is upgrading its tracking chambers, and will have a new
forward calorimeter and extended muon coverage.  It is expected that each
experiment will be able to measure the $W$ boson mass to about 40 MeV.

\section{Conclusion}

The $W$ boson mass has been measured at the Tevatron to a precision of $0.11\%$.
Its value is consistent with the direct LEP2 measurement, the indirect
LEP1/SLC and NuTeV measurements, and the Standard Model.  The D\O\ and CDF 
measurements of diboson production agree with the Standard Model,
and stringent limits have been set on trilinear gauge boson anomalous couplings.
Measurements have been made of the $W$ and $Z$ production cross sections,
the $W$ boson width, and rare $W$ decays, and no disagreement with the
Standard Model has been found.

\section{Acknowledgements}

I would like to acknowledge the generous assistance of my colleagues from the
D\O\ and CDF collaborations who helped in preparing my talk and this paper.
This work was
supported by the Director, Office of Energy Research, Office of High
Energy and Nuclear Physics, Division of High Energy Physics of the
U.S. Department of Energy under Contract DE-AC03-76SF00098.

\end{document}